\documentclass{ws-ijmpd}

\def\eqiaz{\begin{eqnarray*}}
\def\eqfaz{\end{eqnarray*}}
\def\eqia{\begin{eqnarray}}
\def\eqfa{\end{eqnarray}}
\def\btab{\begin{tabular}}
\def\etab{\end{tabular}}
\def\bar{\begin{array}}
\def\ear{\end{array}}
\def\rfr#1{Equation (\ref{#1})}
\def\rfrs#1#2{Equation (\ref{#1})- Equation (\ref{#2})}

\def\Rfrs#1#2{Equation (\ref{#1})- Equation (\ref{#2})}

\def\eqi{\begin{equation}}
\def\eqf{\end{equation}}

\def\lb#1{\label{#1}}
\def\rp#1#2{{#1\over#2}}

\def\dert#1#2{\rp{{ d}{#1}}{{ d}{#2}}}

\begin{document}

\def\nocropmarks{\vskip5pt\phantom{cropmarks}}

\let\trimmarks\nocropmarks

\markboth{Lorenzo Iorio} {A gravitational spin--spin effect}

\catchline{}{}{}

\title{ON SOME GRAVITOMAGNETIC SPIN-SPIN EFFECTS FOR ASTRONOMICAL BODIES}

\author{\footnotesize LORENZO IORIO}

\address{Dipartimento di Fisica dell'Universit$\grave{\rm a}$ di Bari\\
Via Amendola 173, 70126, Bari, Italy}

\maketitle

\begin{abstract}
In this paper we look at the gravitational spin--spin interaction
between macroscopic astronomical bodies. In particular, we
calculate their post--Newtonian orbital effects of order
$\mathcal{O}(c^{-2})$ on the trajectory of a spinning particle
with proper angular momentum ${\bf s}$ moving in the external
gravitomagnetic field generated by a central spinning mass with
proper angular momentum ${\bf J}$. It turns out that, at order
$\mathcal{O}(e)$ in the orbiter's eccentricity, the eccentricity
the pericenter and the mean anomaly rates of the moving particle
are affected by long--term harmonic effects. If, on one hand, they
are undetectable in the Solar System, on the other, maybe that in
an astrophysical context like that of the binary millisecond
pulsars there will be some hopes of measuring them in the future.
\end{abstract}

\section{Introduction}
In recent years the topic of the general relativistic
post--Newtonian effects of  order $\mathcal{O}(c^{-2})$ on various
features of the motion of spinning particles freely orbiting a
central astronomical body has received great attention both
theoretically and experimentally. More precisely, when a spinning
particle moves in an external gravitational field one has to
describe both its spin precession and the influence of the spin on
its trajectory$^1$.

The geodetic, or De Sitter, precession$^2$ refers to the coupling
of the static, gravitoelectric part of the gravitational field due
to the Schwarzschild metric generated by a central, non--rotating
object to the spin of a particle freely orbiting around it. It has
been measured for the Earth--Moon orbit, thought of as a giant
gyroscope, in the gravitational field of the Sun with 1$\%$
accuracy$^3$.  It should be measured for four superconducting
gyroscopes in the gravitational field of the Earth by the
important GP-B mission$^4$ at a claimed relative accuracy level of
$2\times 10^{-5}$. Finally, it might be possible to detect it also
in some binary pulsar systems$^5$.

The Lense--Thirring drag of the inertial frames$^6$ is an effect
due to the stationary, gravitomagnetic part of the gravitational
field of a rotating central mass with proper angular momentum
${\bf J}$ on the geodesic path of a freely falling test particle,
i.e. considered to be not spinning itself\footnote{Such effect
could be thought of as a spin--orbit interaction between the spin
of the central object and the orbital angular momentum of the test
body. For some spin--orbit effects induced by the rotation of the
Sun on the orbital angular momentum of the Earth--Moon system see
reference$^7$.}. In 1998 the first evidence of the Lense--Thirring
effect in the gravitational field of the Earth has been
reported$^8$ with a claimed accuracy of almost $20\%$. It is based
on the analysis of the laser--ranging data of the LAGEOS and
LAGEOS II geodetic satellites. The launch of the proposed
LAGEOS--like LARES satellite$^9$ could allow to measure such
effect with an accuracy probably better than $1\%$. Another
interesting gravitomagnetic effect of order $\mathcal{O}(c^{-2})$
on the orbit of a test particle has been recently derived in
reference$^{10}$; it is due to the temporal variability of the
Earth's angular momentum. Unfortunately, it is too small to be
detected with Satellite Laser Ranging.

The spin of the central object affects also the spin of  a
particle freely orbiting it in a way discovered by Schiff$^{11}$
in 1959. The detection of this subtle precessional effect, in
addition to the geodetic precession, is one of the most important
goals of the GP-B mission$^4$; the claimed accuracy amounts to
$1\%$.

In this paper we are interested in looking for some orbital
effects due to the spin--spin gravitational coupling on the
geodesic path of a spinning extended particle with mass $m$ and
proper angular momentum ${\bf s}$ freely orbiting around a central
body of mass $M$ and proper angular momentum ${\bf J}$. The
gravitational spin--spin coupling in the quantum mechanical domain
has been treated in references$^{12,\ 13}$.

The paper is organized as follows. In Section 2 we derive the
gravitational Stern--Gerlach force with some simplifying
assumptions. In Section 3 we work out the long--term orbital
effects of such interaction on the Keplerian orbital elements of
the moving particle. In Section 4  we look at the Sun--Mercury
system and to PSR B1259-63 and PSR B1913+16 in order to see if the
predicted effect could be measured. Section 5 is devoted to the
conclusions.
\section{The gravitomagnetic Stern--Gerlach force}
In the context of the linearized gravitoelectromagnetism it turns
out that a gravitomagnetic dipole moment for a gyroscope of spin
${\bf s}$ is$^{12}$ \eqi {\overrightarrow{\mu}}_{\rm g}
=-\frac{\bf s}{c}\eqf and the energy of interaction with an
external  gravitomagnetic field ${\bf B }_{\rm g}$ is \eqi H=
-{\overrightarrow{\mu}}_{\rm g}\cdot {\bf B }_{\rm g}=\frac{{\bf
s}\cdot {\bf B }_{\rm g}}{c}\lb{pot}.\eqf Since, in general, the
gravitomagnetic field is position--dependent, a Stern--Gerlach
force arises from \rfr{pot} \eqi {\bf F}=-\nabla H.\lb{forc}\eqf

Let us consider a central spherical rotating body as source of the
gravitomagnetic field, so that \eqi {\bf B }_{\rm g}=-\frac{GJ}{c
r^3}\left[\hat{J}-3\left(\hat{J}\cdot\hat{r}\right)
\hat{r}\right],\eqf where $G$ is the Newtonian gravitational
constant, $\hat{J}$ is the unit vector along the proper angular
momentum of the central body and $\hat{r}$ is the unit position
vector. By assuming ${\bf J}=J\hat{z}$ in an inertial frame with
the $\{x,\ y\}$ plane coinciding with the equatorial plane of the
central mass, \rfr{forc} becomes
  \begin{eqnarray}
    F_x &=& \frac{3GJ}{c^2 r^5}\left[s_x\left(\frac{5x^2 z}{r^2}-z\right)+s_y\left(\frac{5xyz}{r^2}\right)
    +s_z\left(\frac{5xz^2}{r^2}-x\right)\right],\lb{uno}\\
    F_y &=& \frac{3GJ}{c^2 r^5}\left[s_x\left(\frac{5xyz}{r^2}\right)+s_y\left(\frac{5y^2 z}{r^2}-z\right)
    +s_z\left(\frac{5yz^2}{r^2}-y\right)\right],\\
    F_z &=& \frac{3GJ}{c^2 r^5}\left[s_x\left(\frac{5xz^2}{r^2}-x\right)+s_y\left(\frac{5yz^2}{r^2}-y\right)
    +s_z\left(\frac{5z^3}{r^2}-3z\right)\right].\lb{tre}
  \end{eqnarray}
\Rfrs{uno}{tre} agree with the expression \eqi {\bf F}=
\frac{3GJ}{c^2 r^4}\left\{\left[5({\bf
s}\cdot\hat{r})(\hat{J}\cdot\hat{r})-{\bf s
}\cdot\hat{J}\right]\hat{r}-({\bf
s}\cdot\hat{r})\hat{J}-(\hat{J}\cdot\hat{r}){\bf s }\right\}\eqf
of Equation (22) in reference$^{12}$ and with the first term of
Equation (2) in reference$^1$.
\section{The orbital effects}
Now, let us work out the long--term orbital effects on the orbit
of a spinning particle. We will consider it as an extended
spherical body\footnote{It should be noticed that, according to
the authors of reference$^{14}$, the straightforward extension of
the spin--spin gravitational interaction for elementary particles
to macroscopic extended rotating bodies should not be justified.
Then, they follow a phenomenological approach not based a priori
on General Relativity. } of radius $l$ and (slowly) spinning with
an angular velocity $\alpha$, so that $s=\frac{2}{5}ml^2\alpha$.
We will consider the gravitomagnetic spin--spin force of
\rfrs{uno}{tre} as a small perturbation of the Keplerian monopole
and we will adopt the Gauss perturbing equations \begin{eqnarray}
\dert{a}{t} & = & \rp{2}{n\sqrt{1-e^2}}\left[Re\sin f+T\rp{p}{r}\right],\lb{smax}\\
\dert{e}{t} & = & \rp{\sqrt{1-e^2}}{na}\left\{R\sin f+T\left[\cos f+\rp{1}{e}\left(1-\rp{r}{a}\right)\right]\right\},\\
\dert{i}{t} & = & \rp{1}{na\sqrt{1-e^2}}N\rp{r}{a}\cos (\omega+f),\lb{inclinazione}\\
\dert{\Omega}{t} & = & \rp{1}{na\sin i\sqrt{1-e^2}}N\rp{r}{a}\sin (\omega+f),\lb{nodus}\\
\dert{\omega}{t} & = & -\cos i\dert{\Omega}{t}+\rp{\sqrt{1-e^2}}{nae}\left[-R\cos f+
T\left(1+\rp{r}{p}\right)\sin f\right],\lb{perigeo}\\
\dert{\mathcal{M}}{t} & = & n
-\rp{2}{na}R\rp{r}{a}-\sqrt{1-e^2}\left(\dert{\omega}{t}+\cos
i\dert{\Omega}{t}\right),\lb{manom}
\end{eqnarray}
 where $R,\ T,\ N$ are the projections of the perturbing acceleration onto
 the radial $\hat{r}$, transverse $\hat{t}$ and out--of--plane $\hat{n}$ directions of an orthonormal frame
 comoving with the orbiter, $a,\ e,\ i,\ \Omega,\ \omega$ and $\mathcal{M}$ are the
orbiter's semimajor axis, eccentricity, inclination, longitude of
the ascending node, argument of perigee and mean anomaly,
respectively. Moreover, $p=a(1-e^2)$, $f$ is the true anomaly and
$n=\sqrt{GMa^{-3}}$ is the Keplerian mean motion.

Now we will derive the components $R,\ T,\ N$ from
\rfrs{uno}{tre}. In order to simplify the problem we will assume
to consider only equatorial orbits. Moreover, we will further
assume that the proper angular momentums of the two body are
aligned, i.e. ${\bf s}=s\hat{k}$ as well. This is a situation
quite common, e.g., with the planets and their moons in the Solar
System to which our attention will be drawn. Indeed, even if of
order $\mathcal{O}(c^{-2})$, the investigated effects are very
small due to the dependence on $r^{-4}$, as can be noticed from
\rfrs{uno}{tre} and by considering that $x,\ y,\ z$ are all
proportional to $r$; so, only astronomical bodies could yield a
chance for measuring them.

Let us make the $x$ axis coincide with the line of the nodes: in
this case
\begin{eqnarray}
\hat{x}&=& \cos(\omega+f)\hat{r}+\cos i \sin(\omega+f)\hat{t}+\sin
i\sin(\omega+f)\hat{n},\\\lb{kli}
\hat{y}&=&
-\sin(\omega+f)\hat{r}+\cos i \cos(\omega+f)\hat{t}+\sin
i\cos(\omega+f)\hat{n},\\
\hat{z}&=& -\sin i\ \hat{t}+\cos i\ \hat{n},
\end{eqnarray}
and
\begin{eqnarray}
x &=& r\cos(\omega+f),\\
y &=& r\cos i\sin(\omega+f),\\
z &=& r\sin i\sin(\omega+f).\lb{klo}
\end{eqnarray}

With \rfrs{kli}{klo}, \rfrs{uno}{tre} become, in the case of
equatorial orbits
\begin{eqnarray}
R&=&-\rp{3GJ\overline{s}}{c^2 r^4}\cos\left[2(\omega+f)\right],\lb{Uno}\\
T&=&-\rp{3GJ\overline{s}}{c^2 r^4}\sin\left[2(\omega+f)\right],\\
N&=0&\lb{Tre}
\end{eqnarray}
where $\overline{s}=\rp{s}{m}=\rp{2}{5}l^2\alpha$. From \rfr{Tre}
and \rfrs{inclinazione}{nodus} it can be noticed that the
inclination and the node, which, on the other hand, is not defined
for equatorial orbits, are not affected by the gravitomagnetic
spin--spin force.

By calculating \rfrs{Uno}{Tre} on the Keplerian unperturbed orbit
\eqi r_{\rm Kep}=\rp{a(1-e^2)}{1+e\cos f},\eqf neglecting the
terms of order $\mathcal{O}(e^2)$, inserting the obtained result
into \rfrs{smax}{manom} and averaging over an orbital
revolution\footnote{In the averaging process we consider the
vector $\overline{s}$ as constant, i.e. we neglect the De Sitter
and Lense--Thirring precessions of the spin of the orbiting
particle.} by means of \eqi dt=\rp{(1-e^2)^{\rp{3}{2}}}{n(1+e\cos
f)^2}df,\eqf we obtain that only the eccentricity, the perigee and
the mean anomaly experience long--term, harmonic perturbations
with half the period of the perigee.
\begin{eqnarray}
\left\langle\dert{e}{t}\right\rangle
&=&-\rp{3GJ\overline{s}e}{4c^2 a^5 (1-e^2)^2}\sin 2\omega,\lb{spinsemimaj}\\
\left\langle\dert{\omega}{t}\right\rangle
&=&-\rp{15GJ\overline{s}}{4c^2 a^4 (1-e^2)^2}\cos 2\omega,\lb{spinperig}\\
\left\langle\dert{{\mathcal{M}}}{t}\right\rangle
&=&\rp{15GJ\overline{s}}{4c^2 a^4 (1-e^2)^{\rp{3}{2}}}\cos
2\omega.\lb{spinanom}
\end{eqnarray}
\section{Is it possible to measure the gravitomagnetic spin--spin force?}
\subsection{A Solar System scenario}
In order to measure the small effects predicted by
\rfrs{spinsemimaj}{spinperig} the most suitable system seems to be
the Sun--Mercury one. In Table 1 we quote the relevant orbital
parameters of it. For them we have used the reference$^{15}$.
\begin{table}[htbp]
\ttbl{30pc}{Orbital parameters for the Sun and Mercury}
{\begin{tabular}{lccc}\\
\multicolumn{4}{c}{}\\[6pt]\hline
Parameter  & Description & Value & Units \\ \hline
$M_{\odot}$ & Mass of the Sun & $1.9891\times 10^{33}$ & g \\
$R_{\odot}$ & Mean equatorial radius of the Sun & $6.96\times 10^{10}$& cm \\
$\varepsilon_{\odot}$ & Tilt of the Sun's spin axis to the ecliptic & 7.25 & deg\\
$T_{\odot}$ & Sidereal rotational period of the Sun & $24.65$ & days\\
A.U. & Astronomical Unit & $1.4959787066\times 10^{13}$ & cm \\
$M_{\rm m}$ & Mass of Mercury & $3.302\times 10^{26}$ & g \\
$R_{\rm m}$ & Mean equatorial radius of Mercury & $2.4398\times
10^{8}$ & cm\\
$a_{\rm m}$ & Semimajor axis of Mercury & $0.3870$ & A.U. \\
$e_{\rm m}$ & Eccentricity of Mercury & 0.2505& -\\
$i_{\rm m}$ & Inclination of Mercury's orbit to the ecliptic & 7.0048 & deg\\
$\varepsilon_{\rm m}$ & Tilt of Mercury's spin axis to the ecliptic & 0.01 & deg\\
$T_{\rm m}$ & Sidereal rotational period of Mercury & 58.646 &
days\\
$P_{\rm m}$ &Orbital period of Mercury & 0.2408 & years\\
$G$ & Newtonian constant of gravitation &
$6.67259\times 10^{-8}$ &
g$^{-1}$ cm$^3$ s$^{-2}$\\
$c$ & Speed of light in vacuum & $2.9979\times 10^{10}$ & cm
s$^{-1}$ \\
\hline
\end{tabular}}
\end{table}
From Table 1 it can be inferred that, if we choose the ecliptic
plane as $\{x,\ y\}$ plane, the assumptions made by us are
satisfied rather well for the Sun and Mercury. Indeed, the Sun's
spin axis is only slightly tilted to the vertical with respect to
the ecliptic plane, the inclination of Mercury's orbit to it
amounts to few degrees and the spin axis of Mercury is vertical,
so that it can be considered aligned to that of the Sun, at least
to an approximate extent. The values of Table 1 yield for the
Mercury's perihelion shift\eqi
\left\langle\dert{\omega}{t}\right\rangle_{\rm Mercury} =
-(7.6\times 10^{-10}\ {\rm mas/y} )\times \sin 2\omega.
\lb{spinperigmer} \eqf Unfortunately, this is a value which is
completely undetectable.
\subsection{A pulsar scenario}
Let us try to see if the predicted effects could be measured in an
astrophysical context. The binary systems in which at least one
member is a pulsar neutron star seem to be the optimal choice due
to the large values of the eccentricity of the orbits of a not
negligible fraction of them (70 out of more than 1300 known, i.e.
about 5$\%$). Moreover, the rapid rotations of the pulsars and the
relatively small separations between the members of such systems
could be helpful.

As can be inferred from Table 2, obtained from the
data\footnote{The value of the pulsar radius $r$ is an estimate
based on the first calculations by Oppenheimer and
Volkoff$^{16}$.} in reference$^{17}$, the PSR B1259-63 system
seems to be a good choice (See, e.g.,
http://www.jb.man.ac.uk/$\sim$ pulsar/ for a quick outlook of the
discovered pulsars and their relevant orbital parameters). Of
course, in that case our calculations should be considered as very
preliminary order--of--magnitude estimates, just to understand if
looking at pulsars more thoroughly and carefully could be
fruitfully or not.
\begin{table}[htbp] \ttbl{30pc}{Orbital parameters for the PSR
B1259-63/SS 2883 binary system}
{\begin{tabular}{lccc}\\
\multicolumn{4}{c}{}\\[6pt]\hline
Parameter  & Description & Value & Units \\ \hline
$M_{\rm c}$ &
Mass of
the companion & 10 & $M_{\odot}$\\
 $R_{\rm c}$ & Equatorial radius of the companion & 6 & $R_{\odot}$\\
$v_{\rm max}$ & Break--up velocity of the companion & 480 & km
s$^{-1}$\\
$m$ & Mass of PSR B1259--63 & 1.4 & $M_{\odot}$\\
$r$ & Radius of PSR B1259--63 & 10 & km\\
$T$ & Period of rotation of PSR B1259--63 & 47.7620537 & ms\\
$P$ & Period of revolution of PSR B1259--63 & 1,236.7238 & days\\
$i$ & Inclination of PSR B1259--63 & 36 (144) & deg\\
$e$ & Eccentricity of PSR B1259--63 & 0.86990 & --\\
$x=\rp{a\sin i}{c}$ & Projected semimajor axis of PSR B1259--63 &
1296.4 & s\\
\hline
\end{tabular}}
\end{table}
In the case of PSR B1259-63 the companion is SS 2883, a
10th--magnitude star of spectral class B2e, almost 10 times more
massive than the pulsar which orbits it in 3.4 years along a very
eccentric path. In this case we will assume as $\{x,\ y\}$
reference plane the plane of sky which is  a fixed plane normal to
the line--of--sight. For the definition of the various orbital
parameters of the pulsar systems see reference$^{18}$. It should
be noticed that in applying our results to PSR B1259-63 the
following $caveat$ hold. Neither the orbit of the pulsar can be
considered equatorial\footnote{Notice also that from pulsar orbit
data reductions the inclination $i$ cannot be determined
unambiguously.}, nor the spin ${\bf J}$ of SS 2883 is
vertical$^{17}$. Moreover, we can calculate $s$ for the pulsar,
but we cannot say anything about its direction. On other hand, the
value can be inferred for $J$ is plausible because the stars of
the same type of SS 2883 are well known.

By using the values of Table 2 we obtain for the periastron
advance\eqi \left\langle\dert{\omega}{t}\right\rangle_{\rm PSR\
B1259--63} = -(3.3\times 10^{-13}\ {\rm deg/y})\times\sin 2\omega.
\eqf Notice that the period of the secular precession of the
perigee of PSR B1259--63 amounts to 1.9$\times 10^{6}$
years\footnote{It is mainly due to the classical effects of the
oblateness of SS 2883; the relativistic gravitoelectric Einstein's
precession amounts to only 3$\times 10^{-5} $ deg/y.}, so that
over reasonable observational time spans of few years the
predicted harmonics would be similar to secular trends. According
to Table 1 of reference$^{17}$, the experimental sensitivity
amounts to $10^{-6}$ deg/y for $\langle\dot\omega\rangle$, so that
also in this case the gravitomagnetic spin--spin interaction turns
out to be too small to be detected.

A more favorable situation occurs for the well known binary pulsar
system of PSR B1913+16 in which the companion of the pulsar is
probably another neutron star. The relevant orbital parameters are
in Table 3 and have been taken from reference$^{19}$.
\begin{table}[htbp] \ttbl{30pc}{Orbital parameters for the PSR
B1913+16 binary system}
{\begin{tabular}{lccc}\\
\multicolumn{4}{c}{}\\[6pt]\hline
Parameter  & Description & Value & Units \\ \hline $m_{\rm c}$ &
Mass of the companion & 1.3873 & $M_{\odot}$,\\
$m$ &
Mass of PSR B1913+16 & 1.4411 & $M_{\odot}$,\\
$r$ & Radius of PSR B1913+16 & 10 & km\\
$T$ & Period of rotation of PSR B1913+16 & 59.029997929883 & ms\\
$P$ & Period of revolution of PSR B1913+16 & 3.22997462736$\times
10^{-1}$ & days\\
$i$ & Inclination of PSR B1913+16 & 47 (133) & deg\\
$e$ & Eccentricity of PSR B1913+16 & 0.6171309 & --\\
$x=\rp{a\sin i}{c}$& Projected semimajor axis of PSR B1913+16 &
2.341759&
s\\
\hline
\end{tabular}}
\end{table}
In this case we do not know anything about the period of rotation
of the companion because it neither manifests to us in the visible
nor in the radio regions of electromagnetic spectrum. We will
assume for it $10^{-2}$ s, which is rather reasonable since the
periods of the known pulsars range from $1.56\times 10^{-3}$ s to
$6\times 10^{-1}$ s. An--order of--magnitude
calculation\footnote{The spin of PSR B1913+16 may be affected by
the relativistic geodetic precession$^5$, but since its period
would amount to 360 years it can be considered constant over an
orbital revolution.} with our formulas yields \eqi
\left\langle\dert{\omega}{t}\right\rangle_{\rm PSR\ B1913+16} =
-(4.6\times 10^{-7}\ {\rm deg/y})\times\sin 2\omega. \eqf Notice
that the period of the secular precession of the perigee of PSR
B1913+16 amounts to 90 years\footnote{It is entirely due to the
gravitoelectric Einstein's precession.}, so that over reasonable
observational time spans of few years the predicted harmonics
would be similar to secular trends. Since for PSR B1913+16 the
experimental accuracy in measuring
$\left\langle\dot\omega\right\rangle$ amounts to about $1\times
10^{-5}$ deg/y (see reference$^{20}$), in this case we are not too
far from the the possibility of detecting the predicted effect.
\section{Conclusions}
In this paper, in the linearized approximation of
gravitoelectromagnetism and at order $\mathcal{O}(c^{-2})$,  we
have calculated the influence of the spin ${\bf s }$ of a particle
on its geodesic orbital motion in an external gravitomagnetic
field generated by a rotating body with spin ${\bf J }$. It has
been assumed that the orbit lies in the equatorial plane of the
central object and that the spins are both vertical to it. The
orbital--averaged, long--term effects on the Keplerian orbital
elements of the orbiter have been calculated by neglecting all
terms of order $\mathcal{O}(e^{2})$. It turns out that the
eccentricity, the pericenter and the mean anomaly are affected by
such spin--spin gravitomagnetic interaction by means of harmonic
perturbations with half the period of the pericenter. In order to
see if they are measurable, we have examined two possible
astronomical scenarios: the Sun--Mercury system and a pair of
binary millisecond pulsar systems. It turns out that they are far
too small. Only for the pulsar PSR1913+16 the predicted periastron
rate is not too far from the present experimental sensitivity. The
possible discovery of new, highly eccentric and close binary
pulsar systems, together with notable improvements of the accuracy
in measuring the periastron rates could give some hopes to detect
such tiny effects in future.
\section*{Acknowledgements} I wish to thank H. Lichtenegger for his kind hospitality at IWF in Graz and
B. Mashhoon for the kindly suggested references.

\end{document}